\documentclass[preprint2,longabstract]{aastex}

\renewcommand{\b}[1]{\mbox{\boldmath $#1$}}

\def\cal#1{{\cal #1}}

\def\m@th{\mathsurround=0pt}
\def\n@space{\nulldelimiterspace=0pt \m@th}
\def\biggg#1{{\mbox{$\left#1\vbox to 20.5pt{}\right.\n@space$}}}

\def\beginenum{\begin{enumerate}}
\def\endenum{\end{enumerate}}
\def\bitem{\begin{itemize}}
\def\eitem{\end{itemize}}
\def\bray{\begin{array}}
\def\eray{\end{array}}
\def\begindoc{\begin{document}}
\def\enddoc{\end{document}}
\def\bq{\begin{equation}}
\def\eq{\end{equation}}
\def\bqy{\begin{eqnarray}}
\def\eqy{\end{eqnarray}}
\def\bqyn{\begin{eqnarray*}}
\def\eqyn{\end{eqnarray*}}
\def\bc{\begin{center}}
\def\ec{\end{center}}
\def\bfll{\begin{flushleft}}
\def\efll{\end{flushleft}}
\def\bflr{\begin{flushright}}
\def\eflr{\end{flushright}}
\newcommand{\Avec}{\mbox{\boldmath $A$}}
\newcommand{\Bvec}{\mbox{\boldmath $B$}}
\newcommand{\Evec}{\mbox{\boldmath $E$}}
\newcommand{\Fvec}{\mbox{\boldmath $F$}}
\newcommand{\Gvec}{\mbox{\boldmath $G$}}
\newcommand{\Rvec}{\mbox{\boldmath $R$}}
\newcommand{\Uvec}{\mbox{\boldmath $U$}}
\newcommand{\Vvec}{\mbox{\boldmath $V$}}
\newcommand{\evec}{\mbox{\boldmath $e$}}
\newcommand{\jvec}{\mbox{\boldmath $j$}}
\newcommand{\kvec}{\mbox{\boldmath $k$}}
\newcommand{\nvec}{\mbox{\boldmath $n$}}
\newcommand{\uvec}{\mbox{\boldmath $u$}}
\newcommand{\vvec}{\mbox{\boldmath $v$}}
\newcommand{\wvec}{\mbox{\boldmath $w$}}
\newcommand{\xvec}{\mbox{\boldmath $x$}}
\newcommand{\omegavec}{\mbox{\boldmath $\omega$}}
\newcommand{\Omegavec}{\mbox{\boldmath $\Omega$}}

\usepackage{spr-astr-addons}    

\usepackage{color}

\begin{document}
%
\title{Mechanisms for Multi-Scale Structures in Dense Degenerate Astrophysical Plasmas}

\shorttitle{<Multi-Scale Equilibrium Structures>}
\shortauthors{<Shatashvili, Mahajan \& Berezhiani >}

\author{ N.L. Shatashvili\altaffilmark{1,2}}
\altaffiltext{1}{Andronikashvili Institute of Physics, TSU,
Tbilisi 0177, Georgia} \
\altaffiltext{2}{Department of Physics,
Faculty of Exact and Natural Sciences, Ivane Javakhishvili Tbilisi
State University (TSU), Tbilisi 0179, Georgia}
\author{ S.M. Mahajan\altaffilmark{3}}
\altaffiltext{3}{Institute for Fusion Studies, The University of
Texas at Austin, Austin,Tx 78712} \and
\author{V.I. Berezhiani\altaffilmark{1,4}}
\altaffiltext{1}{Andronikashvili Institute of Physics, TSU,
Tbilisi 0177, Georgia}
\altaffiltext{4}{School of Physics, Free
University of Tbilisi, Georgia} \

\begin{abstract}
Two distinct routes lead to the creation of multi--scale
equilibrium structures in dense degenerate plasmas, often met in
astrophysical conditions. By analyzing an e-p-i plasma consisting
of degenerate electrons and positrons with a small contamination
of mobile classical ions, we show the creation of a new macro
scale $L_{\rm{macro}}$ (controlled by ion concentration). The
temperature and degeneracy enhancement effective inertia of bulk
e-p components also makes the effective skin depths larger (much
larger) than the standard skin depth. The emergence of these
intermediate and macro scales lends immense richness to the
process of structure formation, and vastly increases the channels
for energy transformations. The possible role played by this
mechanism in explaining the existence of large-scale structures in
astrophysical objects with degenerate plasmas, is examined.
\end{abstract}


\maketitle


\section{Introduction}

Dense compressed plasmas, found in astrophysical and cosmological
environments as well as in laboratory experiments investigating
interaction of intense lasers with high density plasma, are
currently of great interest. In dense astrophysical objects like
white and brown dwarfs, neutron stars, magnetars, and cores of
giant planets, extreme conditions lead to high density degenerate
matter \citep{Sturok-1,Compact-1,Chandra1} ;\\
\citep{Chandra2,Sturok-2,Sturok-3,Compact-2,White,Beloborodov}.
Most astrophysical plasmas usually contain ions in addition to
degenerate electrons and positrons. Although there is no concrete
evidence, the magnetospheres of rotating neutron stars are
believed to contain electron-positron plasmas produced in the cusp
regions of the stars due to intense electromagnetic radiation.
Since protons or other ions may exist in such environments,
three-component electron-positron-ion (e-p-i) plasmas can exist in
pulsar magnetospheres \citep{Begelman,Tajima-1},\\
\citep{Tajima-2}.

For typical dense plasmas, composed of ions,  electrons,
positrons, and/or holes (in the context of semiconductors), the
lighter species (electrons, positrons, holes) are degenerate while
the more massive ions, often, remain nondegenerate (classical).

It has been known that the nonlinear phenomena in e-p plasmas
develop differently from their counterparts in the usual
electron--ion system. The positron component could have a variety
of origins: 1) positrons can be created in the interstellar medium
due to the interaction of atoms and cosmic ray nuclei, 2) they can
be introduced in a Tokamak e-i plasma by injecting bursts of
neutral positronium atoms ($e^+\,e^-$), which are then ionized by
plasma; the annihilation time of positron in the plasma is long
compared to typical particle confinement time \citep{Mamun}. The
annihilation, which takes place in the interaction of matter
(electrons) and anti-matter (positrons), usually occurs at much
longer characteristic time scales compared with the time in which
the collective interaction between the charged particles takes
place \citep[and references therein]{degenerate}.

In the extremely low temperature three component e-p-i plasmas,
studied in the context of of pulsar magnetospheres in \citep{EPI-1,EPI-3} ;\\
\citep{EPI-2,EPI-5}, the \\
de Broglie wavelength of the charge carriers can be comparable to
the dimension of the systems. Such ultracold e-p-i plasma behaves
like a Fermi gas and quantum mechanical effects are expected to
play a significant role in their linear and nonlinear dynamics.

Pursuing the consequences of degeneracy in a multi-component
plasma, it will be interesting to explore multiple-scale behavior
accessible to such systems.  An obvious goal will be to
investigate if degeneracy can affect, for example, the dynamics of
the star collapse when its multi-species atmosphere begins to
contract. Analysis of multi-scale behavior can also be a guide in
predicting various phenomena in the pre-compact era, or during the
life of compact objects.

The principal determinant of degeneracy, the density, varies over
many orders of magnitude in astro/cosmic settings. The rest frame
e-p density near the pulsar surface is believed to be $n\geq
10^{11}\,cm^{-3}$ \citep{Misha}, while in the MeV epoch of the
early Universe , it  can be as high as $n=10^{32}\,cm^{-3}$
\citep{Weinberg}. Intense e-p pair creation takes place during the
process of gravitational collapse of massive stars
\citep{Stenflo}. It is argued that the gravitational collapse of
the massive stars may lead to  charge separation with the field
strength exceeding the Schwinger limit resulting in e-p pair
plasma creation with estimated density to be $n=10^{34}\,cm^{-3}$
\ \citep{ruffini}. Superdense e-p plasma may also exist in GRB
sources [$n=(10^{30}-10^{37})\,cm^{-3}$] \citep{aksenov}.

Dense electron-positron plasma can be produced in laboratory
conditions as well. Indeed, the modern petawatt lasers systems are
already capable of producing ultrashort pulses with  focal
intensities \ $I=2\times 10^{22}\,W/cm^{2}$ \ \citep{Yanovski}.
Pulses of even higher intensities exceeding \
$I=10^{26}\,W/cm^{2}$ \ are likely to be available soon in lab or
in the Lorentz boosted frames \citep{Dunne-1,Dunne-2} ;\\
\citep{Dunne-3}. Interaction of such pulses with gaseous or solid
targets could lead to the generation of optically thin e-p plasmas
($n\sim 10^{23}-10^{28}\,cm^{-3}$) denser than solid state
systems\ \citep{Shukla-Eliasson}.

In the highly compressed degenerate Fermi state (average
inter-particle distance smaller than the thermal de Broglie
wavelength), mutual interactions of the plasma particles become
unimportant, and plasma becomes more ideal as the density
increases \\
\citep{Landau}. A plasma may be considered cold (functionally zero
temperature) if the thermal energy of the particles is much lower
than the Fermi energy no matter how high the temperature really is
\citep{Russo-1,Russo-2}. The Fermi energy of degenerate electrons
(positrons) is \ \ $\epsilon _{F}=m_{e}c^{2}\,\left[ \left(
1+R^{2}\right) ^{1/2}-1\right] $ , where \ $R=p_{F}/m_{e}c$ \ and
\ $p_{F}$ \  is the Fermi momentum determined by the rest-frame
density \ $p_{F}=m_{e}c\,\left( n/n_{c}\right)^{1/3}$ \ .
Densities are normalized to the critical number-density
$n_{c}=5.9\times 10^{29}\,cm^{-3}$ \ \citep{Akbari}.

The fluid models are frequently applied to study the large scale
dynamics of relativistic multi-species plasmas
\citep{Gedalin-1,Gedalin-2}. Among such investigations the studies
on {\it relaxed (equilibrium) states} have attracted considerable
attention \citep{relaxed}. Constrained minimization of fluid
energy with appropriate helicity invariants has provided a variety
of extremely interesting equilibrium configurations that have been
exploited and found useful for understanding laboratory as well as
astrophysical plasma systems [see e.g.
\citep{relaxed,woltjer,taylor,sudan},\\
\citep{dewar,dewar2} and references therein]. Two particularly
simple manifestations of this class of equilibria -- {\it Beltrami
states} -- are: 1) The single Beltrami state, $ \nabla \times {\bf
B}=\alpha {\b B}$, discussed by Woltjer and Taylor
\citep{woltjer,taylor} in the context of force free single fluid
magnetohydrodynamics (MHD), and 2) a more general Double Beltrami
State accessible  to Hall MHD -- a ``two-fluid' system of ions and
inertia-less electrons \citep{DB}; the latter has been
investigated, in depth, by Mahajan and co-workers \
\citep{MY,mmns-1},\\
\citep{osym,sm}, \\
\citep{mnsy,msms}, \\
\citep{DBwaves1,DBwaves2}. The content of the Beltrami conditions
(derived by constrained minimization) is the alignment of a
general ``flow" with its vorticity. This, in turn, forces the
total energy density of the system to distribute homogeneously (so
called {\it Bernoulli condition}). The combined Beltrami-Bernoulli
conditions define an equilibrium state that fits the notion of
{\it relaxed state}, and constitutes a nontrivial helicity-bearing
state. The characteristic number of a state is determined by the
number of independent single Beltrami systems needed to construct
it. For adequate description these states were named {\it
Beltrami-Bernoulli States} (BB)  \citep{MY,mmns-1,BSM_deg}; the
latter ref. was also the first to study the effects of degeneracy
(requiring Fermi-Dirac statistics) on BB states.

This paper, though worked out on the lines of \citep{BSM_deg},
registers a major departure leading to the most important result
of this article -- by studying the BB states in an e-p-i (small
dynamic ion contamination added to a primarily e-p plasma), we
will demonstrate the creation of a new macroscopic length scale \
$L_{\rm{macro}}$ \ lying between the system size and the
relatively small intrinsic scales (measured by the skin depths) of
the system.

The new BB equilibrium is defined by: two relativistic Beltrami
conditions (one for each dynamic degenerate species), one
non-relativisc Beltrami condition for ion fluid, an appropriate
Bernouli condition, and Ampere's law to close the set. This set of
equations will lead to what may be called a {\it quadruple
Beltrami system} [for multi-Beltrami systems see \citep{Multi-B}].
The ions, though a small mobile component, play an essential role,
they create an asymmetry in the electron-positron dynamics (to
maintain charge neutrality, there is a larger concentration of
electrons than positrons) and that asymmetry introduces a new and
very important dynamical scale. This scale, though present in a
classical non-degenerate plasma, turns out to be degeneracy
dependent and could be vastly different from its classical
counterpart. The significance of this scale in understanding the
physics of relevant systems will be explored. Presence of mobile
ions leads to "effective mass" asymmetry in electron and positron
fluids, which, coupled with degeneracy-induced inertia, manifests
in the existence of Quadruple Beltrami fields. Illustrative
examples and application for concrete astrophysical systems will
be suggested.

\section{Model Equations}

Charge neutrality in an e-p-i plasma of degenerate electrons
($-$), positrons ($+$) and a small mobile ion component, forces
the following density relationships
\begin{equation}
N_0^-=N_0^+ + N_{i0} \qquad \Longrightarrow \qquad
\frac{N_0^+}{N_0^-} = 1-\alpha  \  \label{B-eq}
\end{equation}
\[
\qquad {\rm with} \qquad \alpha = \frac{N_{i0}}{N_0^-} \ ,
\]
where \ $\alpha$ \ labels the excess electron content.  In this
paper we will  study only the \ $\alpha\ll 1$ \ limit that we
believe is the most pertinent for astro/cosmic settings.

The equation for Ion dynamics is standard \\
\citep{BSM_deg}. The {\it e(p)} dynamics will be described by the
relativistic degenerate fluid equations \citep[and references
therein]{BSM_deg,degenerate}: the continuity

\begin{equation}
\frac{\partial N^{\pm}}{\partial t} + \nabla \cdot ({N^{\pm}{\bf
V}_{\pm}})=0 \ , \label{Cont}
\end{equation}

\noindent and the equation of motion

\begin{equation}
\frac{\partial}{\partial t}\left(G^{\pm}{\bf p}_{\pm}\right)+
m_{\pm}c^{2}\, {\bf \nabla}\left(G^{\pm}\gamma_{\pm}\right) =
q_{\pm}\,{\bf E} \ + \ {\bf V_{\pm}\times \Omega_{\pm}} \label{B6}
\end{equation}

\noindent where  \ ${\bf p}_{\pm}= \gamma_{\pm} m_{\pm}{\bf
V}_{\pm}$ \ is the hydrodynamic momentum, \ $n^{\pm} =
N^{\pm}/\gamma_{\pm}$ \ is the rest-frame particle density \
($N^{\pm}$ \  denotes laboratory frame number density), \
$q_{\pm}(m_{\pm})$ \ is the charge (mass) of the positron
(electron) fluid element, \ ${\bf V}_{\pm}$ \ is the fluid
velocity, and $\gamma_{\pm} =\left(
1-V_{\pm}^{2}/c^{2}\right)^{-1/2}$ \ . Notice that the degeneracy
effects manifest through the "effective mass" factor  \ $G^{\pm} =
w_{\pm}/n^{\pm}m_{\pm}c^{2}$ , \ where \ $w_{\pm}$ \ is an
enthalpy per unit volume. The general expression for enthalpy \
$w_{\pm}$ \ for arbitrary density and temperature (for a plasma
described by local Dirac-Juttner equilibrium distribution
function) can be found in \citep{Russo-2}. For a fully degenerate
(strongly degenerate) {\it e(p)} plasma, however, this very
tedious expression smoothly transfers to the one with just density
dependence, i.e, $w_{\pm}\equiv w_{\pm}(n)$ \citep{BSM_deg}. In
fact \ $w_{\pm}/n^{\pm}m_{e}c^{2}=\left(
1+(R^{\pm})^{2}\right)^{1/2}$, where \ $R^{\pm}$ [$=
p_{F\,\pm}/m_{\pm}c$ \ with \ $p_{F\,\pm}$ \ being the Fermi
momentum] have been defined earlier. The mass factor, then, is
simply determined by the plasma rest frame density,
 \ $G^{\pm}=[ 1+(n^{\pm}/n_{c})^{2/3}]^{1/2}$ \ for arbitrary \ $n_{\pm}/n_c$ .

On taking the {\it curl} of these equations, one can cast them
into an ideal vortex dynamics (Mahajan 2003 and references
therein)
\begin{equation}
\frac{\partial}{\partial t}\,{\bf \Omega}_{\pm}={\bf
\nabla\times}\left( {\bf V_{\pm}\times\Omega_{\pm}}\right) \ ,
\label{B7}
\end{equation}
in terms of the generalized (canonical) vorticities \  ${\bf
\Omega_{\pm}=}( q_{\pm}/c ) {\bf B+\nabla\times}\left( G^{\pm}{\bf
p}_{\pm}\right)$ \ .

\bigskip

It is, perhaps, the right juncture to re-emphasize that the so
called plasma approximation for a degenerate {\it e(p)} assembly
is valid if their average kinetic energy ($\sim \epsilon_F^{\pm}$)
is larger than the interaction energy ($\sim
e^2(n_0^{\pm})^{1/3}$). This condition is fulfilled for a
sufficiently dense fluid when $n_0^{\pm}\gg
(2m_-\,e^2/(3\pi^2)^{2/3}{\hbar}^2)^3=6.3\cdot 10^{22}\,cm^{-3}$;
such a condition would imply \ $R^{\pm}\gg 4.76\cdot 10^{-3}$ \
\citep{degenerate}.

\bigskip

The low frequency dynamics is, now, closed with Ampere's law
\begin{equation}
\nabla \times {\bf B} =\frac{4\pi e }{c}\, \left[ (1-\alpha)\,N^+
\,{\bf V}_+ - N^-\,{\bf V}_-) +\alpha \,N_i\,{\bf V}_i\right] ,
\label{B-Amp1}
\end{equation}
another relation between \ ${\bf V}_i \ , {\bf V}_{\pm}$ \ and \
${\bf B}$. Notice that the small static/mobile ion population,
represented by \ $\alpha$ \ and \ ${\bf V}_i$, creates an
asymmetry between the currents contributed by the electrons and
positrons. This will be the source of a new scale-length that
turns out to be much larger than the intrinsic electron and
positron scale lengths (skin depths).

\bigskip

In this paper, we explore the combined effects of asymmetry and
degeneracy on a special class of e-p-i equilibria known as the
Beltrami-Bernoulli (BB) states. We expect to find  large-scale
structures originating in the ion-induced asymmetry. The e-p-i
plasma system is, in some sense, more advanced and complete than
the electron-ion system studied in a recent paper where it was
shown that the electron degeneracy  transformed BB states may be
pertinent to advance our understanding of the evolution of certain
astrophysical objects \citep{BSM_deg}.

\section{Equilibrium States in relativistic degenerate e-p-i Plasma}

Before we write down the equations for the BB states, it is useful
to express all physical quantities in normalized dimensionless
form. In this paper, the density is normalized to \ $N_0^-$ (the
corresponding rest-frame density is \ ${n_0}^-$); the magnetic
field is normalized to some ambient measure $|{\bf B}_0|$; all
velocities are measured in terms of the corresponding Alfv\'en
speed \  $V_A = V_A^{-} = B_0/\sqrt{8\pi n_0^{-}m^-\,G_0^-}$ ; all
lengths [times] are normalized to the "effective" electron skin
depth \ $\lambda_{\rm{eff}} \, [\lambda_{\rm{eff}}/V_A]$ , where
\begin{equation}
\lambda_{\rm{eff}} \equiv
\lambda_{\rm{eff}}^{-}=\frac{1}{\sqrt{2}} \
\frac{c}{\omega_p^-}=c\,\sqrt{\frac{m^-\,G_0^-}{8\pi n_0^- e^2}} \
; \label{lamb}
\end{equation}
\begin{equation}
G_0^{\pm}(n_0^{\pm}) = [1 + (R_0^{\pm})^2]^{1/2} \qquad {\rm with}
\label{G0}
\end{equation}
\begin{equation}
R_0^{\pm}=\left( \frac{n_0^{\pm}}{n_c}\right)^{1/3}; \qquad
R_0^+=(1-\alpha)^{1/3}\,R_0^-.
\label{R0}
\end{equation}
The intrinsic skin depths, the natural length scales of the
dynamics, are generally much shorter compared to the system size.
For the degenerate electron fluid, the effective mass goes from \
$G_0^-(n_0^-) = 1 + \frac{1}{2}(\frac{n_0^-}{n_c})^{2/3}$ \ in the
non-relativistic limit ($R_0^- \ll 1$ ) to  \ $G_0^-(n_0^-) =
\left(\frac{n_0^-}{n_c}\right)^{1/3}$ in the ultra-relativistic
regime ($ R_0^-\gg 1$ ).

By following the methodology of \citep{pino}, where the BB states
were derived for classical relativistic non-degenerate multi-fluid
plasmas, we obtain the required set of equilibrium equations for
the degenerate system (the primary difference is in the physics of
\ $G_{\pm}$ \citep{BSM_deg}): The Beltrami conditions
\begin{equation}
{\bf B}\pm \nabla \times (G^{\pm}\,\gamma_{\pm}{\bf V}_{\pm}) =
a_{\pm}\,\frac{n^{\pm}}{G^{\pm}}\,(G^{\pm}\,\gamma_{\pm}{\bf
V}_{\pm}) \ , \label{B9}
\end{equation}
aligning the Generalized vorticities along their velocity fields,
and the Bernoulli conditions
\begin{equation}
\nabla (G^{\pm}\,\gamma_{\pm}\ \pm \ \varphi )=0 \quad
\Longrightarrow \label{B8}
\end{equation}
\begin{equation}
\qquad \qquad G^+ \,\gamma_++ G^-\,\gamma_- = const \ .
\label{Bernoulli}
\end{equation}
In the latter, $\varphi \neq 0$ \ due to the asymmetry, but
gravity is ignored.

The separation constants \ $a_{\pm}$ \ are related to the total
system energy, and the {\it generalized helicities}
\begin{equation}
h_{\pm}=\int ( {curl^{-1}{\bf \Omega}_{\pm}) \cdot {\bf
\Omega}_{\pm}\,d {\bf r} } \ . \label{B10}
\end{equation}

This set, coupled with Ion fluid Beltrami Condition:
\begin{equation}
{\bf B} + \zeta \nabla \times {\bf V}_i = \alpha \,a_i n_i{\bf
V}_i , \ \ \ \rm{where} \ \ \ \zeta = \left[G_0^-\,
\frac{m^-}{m_i}\right]^{-1} \label{IBC}
\end{equation}
together with Ampere's law  Eq.(\ref{B-Amp1}) defines the BB
system for an e-p-i plasma -- a degenerate e-p system made
somewhat asymmetric by a small fraction of mobile ions ($\alpha\ll
1 , \ |{\bf V}_i|\ll|{\bf V}_{\pm}| $).

Notice that  there are, in fact, two asymmetry-introducing
mechanisms in the e-p-i system:  different effective inertias for
the positively and negatively charged particles of the bulk
species is one, while the small contamination from mobile ions
($\alpha \neq 0 , {\bf V}_i \neq 0$) constitutes the other. Each
one of these is responsible for creating a net ``current''. The
structure formation mechanism explored in
\citep{structuresPI-1,structuresPI-2} \citep{DB},\\
\citep{mmns-1,mnsy}, \\
\citep{osym}), originates, for instance, in the effective inertia
difference. Asymmetry between the plasma constituents increases
the number of conserved helicities, and eventually translates into
a higher index Beltrami state. Later, we will explicitly show that
for the degenerate e-p-i system, the Beltrami part of BB state is
a {\it Quadruple Beltrami} state. When the second asymmetry
mechanism is neglected ($\alpha \to 0 , {\bf V}_i \to 0$) \, a
{\it Triple Beltrami} State follows \citep{dasgupta,Multi-B}.

It should also be mentioned that  initial asymmetry in densities
($\alpha \neq 0$) can also contribute to an  "effective inertia"
difference in the electron-positron plasma making \ $G^- \neq
G^+$; the index of the Beltrami system, however, is contingent on
the simultaneous presence of  both asymmetries (see Appendix A).

\section {The Quadruple Beltrami system}

An appropriate but tedious manipulation of the set Eq.-s
(\ref{B-Amp1})-(\ref{IBC}),  carried out in Appendices A and B,
leads us to an explicit quadruple Beltrami equation obeyed by the
Ion Fluid Velocity ${\bf V}_i$ (the Beltrami index is measured by
the highest number of \ {\it curl} \ operators). Written
schematically as

\[
\nabla \times \nabla \times \nabla \times \nabla \times {\bf V}_i
\ -\ b_1^{'} \,\nabla \times \nabla \times \nabla \times {\bf V}_i
\ +
\]
\begin{equation}
+ \ b_2^{'} \,\nabla \times \nabla \times {\bf V}_i \
 - \ b_3^{'} \,\nabla \times {\bf V}_i \ + \ b_4^{'}\,{\bf V}_i = \ 0 \ .
\label{V-quadruple}
\end{equation}
Equation (\ref{V-quadruple}) was derived in the incompressible
approximation, and for \  $\gamma_+\sim \gamma_-\equiv 1$.  The
$b^{'}$ coefficients are defined in Appendix B. Incompressibility
assumption is expected to be adequate for outer layers of compact
objects, though, compressibility effects can be significant e.g.
in the atmospheres of pre-compact stars \citep{BSM_deg}. The ion
fluid velocity and the magnetic field are related to the e-p
plasma average bulk fluid velocity
\begin{equation}
\qquad \qquad {\bf V} = \frac{1}{2}\,[(1-\alpha ){\bf V}_+ \ + \
{\bf V}_-] \ , \label{B12}
\end{equation}
through
$$
{\bf V} =
$$
$$
=\eta \left( 2\beta G_0^+\nabla \times \nabla \times {\bf B} -
[a_+(1-\alpha)\beta - a_-]\nabla \times {\bf B}\right)+
$$
\begin{equation}
\qquad + \ \eta \, \left( [1 + (1-\alpha)\,\beta ]\,{\bf B}
\right) - \label{Velocity}
\end{equation}
\[
\qquad -\ \alpha \beta \, \nabla \times {\bf V}_i \ + \
\frac{\alpha}{2}\,[a_+(1-\alpha)\beta - a_-]\,{\bf V}_i
\]
\[
{\rm with} \quad \eta \,\equiv [a_+(1-\alpha )\beta + a_-]^{-1} \
\  {\rm and} \ \ \beta = G_0^-/G_0^+ \ .
\]
Once Eq.(\ref{V-quadruple}) is solved for \ ${\bf V}_i$, the
vector fields \ ${\bf B}$ \ and \ ${\bf V}$ \ can be determined
from Eqs.(\ref{IBC})-Eq.(\ref{Velocity}). We have chosen to work,
here, with the more familiar e-p plasma bulk velocity ${\bf V}$
rather than the normalized momenta ${\bf P}_{\pm} =
G^{\pm}(n^{\pm})\,{\bf V}_{\pm}$. The e-p-i system is symmetric in
$\bf B$, $\bf V$, ${\bf V}_i$ in the sense that either of them
obeys a {\it quadruple curl} equation. The {\it curl curl curl}
(Triple Beltrami) equation and its solutions describing a
compressible degenerate pure e-p plasma are given in Appendix C.

In Appendix B we give some illustrative examples of Quadruple
[Triple] Beltrami states for degenerate e-p-i plasmas interesting
for astrophysical context. Since the effect of compressibility in
degenerate e-i plasma was studied in \citep{BSM_deg} we do not
discuss it here and, instead, concentrate on emphasizing the
effects of asymmetry stemming from the dynamic ion contamination.

\bigskip

The quadruple Beltrami ({\ref{V-quadruple}}) can be factorized as
(details in Appendix B)
\begin{equation}
(curl - \mu_1)(curl - \mu_2)(curl - \mu_3)(curl - \mu_4)\ {\bf
V}_i = 0 \ , \label{Q-curl}
\end{equation}
where \ $\mu_i$-s \ define the coefficients in
Eq.(\ref{V-quadruple}) and are the functions of \ $\alpha , \
\beta , \ n_0^-$ \ and the degeneracy-determined mass factor \
$G_0^{+}$ . The general solution of Eq.(\ref{Q-curl}) is a sum of
four Beltrami fields ${\bf F_k}$ (solutions of Beltrami Equations
$\nabla \times {\bf F}_k = \mu_k {\bf F}$) \ while eigenvalues
($\mu_k $) of the {\it curl} operator are the solutions of the
fourth order equation
\begin{equation}
\qquad \qquad \mu^4-b_1^{'}\,\mu^3 + b_2^{'}\,\mu^2 - b_3^{'}\,\mu
+ b_4^{'} = 0 \ .
\label{mu-eq}
\end{equation}

An examination of the various $b^{'}$ coefficients of
(\ref{mu-eq}), displayed in detail in (\ref{QB-2},\ref{QB-3}) for
the most relevant limit \ $\alpha \ll 1$ , reveal the most
interesting and important result of this enquiry. Though the
inverse scales, determined by \ $b_1^{'}$, $b_2^{'}$, and
$b_3^{'}$ , do get somewhat modified by $\alpha \ll 1$
corrections, it is the inverse scale associated with \ $b_4^{'}$ \
that is most profoundly affected; being proportional to $\alpha$,
it tends to become small, i.e, {\it the corresponding scale length
becomes large} as $\alpha$ approaches zero; the scale length
becomes strictly infinite for \ $\alpha = 0$, and disappears
reducing (\ref{mu-eq}) to a triple Beltrami system. Thus the ion
contamination-induced asymmetry may lead to the formation of
macroscopic structures through creating an intermediate/large
length scale, much larger than the intrinsic scale skin depths,
and less than the system size. The possible significance and
importance of this somewhat natural mechanism (a small ion
contamination is rather natural) for creating Macro-structures in
astrophysical objects, could hardly be overstressed. It is
important to note that this mechanism operates for all levels of
degeneracy (the range of \ $R_0^-$ \ was irrelevant).

This new macroscopic scale can be ``determined" by dominant
balance arguments; as the scale gets larger, \ $|\nabla|$ \ gets
smaller, and the dominant balance will be between the last terms
of (\ref{mu-eq}), yielding [we remind the reader, that all lenghts
are normalized to the \ $\lambda_{\rm{eff}}$ , and \ $\zeta \gg 1
$ \ even for ultra-relativistic case]
\begin{equation}
\qquad \qquad L_{\rm{macro}}={\frac{|b_3^{'}|}{|b_4^{'}|}} =
\frac{A}{\alpha} \label{L_macro}
\end{equation}
where
\begin{equation}
A = \zeta \,\frac{| (a_+ - a_-)[1 - \frac{\alpha}{\zeta}\,(G_0^+)]
+ \frac{\alpha}{\zeta}\,a_i\,[(G_0^+)-a_+a_-]|}{|a_i\,(a_+ - a_-)
-a_+a_-| } \label{A}
\end{equation}
is a somewhat complicated function of the plasma parameters.

Assuming that the densities of the e-p-i plasmas of interest are
such that \ $\alpha G_0^+/\zeta = \alpha \,\beta (G_0^+)^2
\frac{m^-}{m_i} \leq \alpha \ll 1 $ \
[{\it e(p)} component density range is within \ $(10^{25} -
10^{32})\,cm^{-3}$] , we can simplify \ $A$ \ when both \ $a_+\ll
a_i $ \ and \ $a_-\ll a_i$.

\noindent (i) When \ $a_+$ \ and \ $a_-$ \ are not equal the
simplified expression for dimensionless
\begin{equation}
\qquad \qquad L_{\rm{macro}} \sim \frac{\zeta}{\alpha }\
\left|{\frac{1}{a_i} +
\frac{\alpha}{\zeta}\,\frac{(G_0^+)-a_+a_-}{a_+ - a_-}}\right|
\label{l1}
\end{equation}
for \ $a_i \leq \zeta$ \ satisfies \ $L_{\rm{macro}} \gg 1$ .
While

\noindent (ii) when \ $a_+ \sim a_- = a \neq (G_0^+)^{1/2}$
\ 
\begin{equation}
\qquad \qquad L_{\rm{macro}} \sim
\frac{a_i}{a^2}\,|(G_0^+)-a^2|\gg 1
\label{l2}
\end{equation}
for all \ $a_i \gg a$ .

Without ion contamination ($\alpha = 0$), the degenerate e-p
system is still capable of creating length scales larger than the
non-relativistic skin depths through the degeneracy-enhanced
inertia of the light particles.  Notice that even with equal
effective masses ($G^- = G^+ \simeq G(n)$ at equal
electron-positron temperature), inertia change due to degeneracy
can cause asymmetry in {\it e(p)} fluids [see \citep{Multi-B}].

\bigskip

This, perhaps, is the right juncture to summarize the scale
hierarchy encountered in this paper:

1) For a pure electron-positron plasma, the equilibrium is {\it
triple Beltrami} with the following fundamental three scales;
system size $L$, and the two intrinsic scales (electron and
positron skin depths).

2) The e-p skin depths, microscopic in a non degenerate plasma,
can become much larger due to degeneracy effects and could be
classified as meso-scales, $l_{\rm{meso}}$ [see Appendix C,
Eq.(\ref{B18})].

3) When a dynamic low density ion species is added, the
equilibrium becomes {\it quadruple Bertrami} with a new additional
scale, $L_{\rm{macro}}$. Although the exact magnitude of this
scale is complicated [Eq. (\ref{l1})], its origin is entirely due
to the ion contamination; this scale disappears as the ion
concentration \ $\alpha $ \ goes to zero. Both the larger ion mass
and low density contribute towards boosting \ $L_{\rm{macro}}$ \
[see Appendix A].

4) The meso-scale $l_{\rm{meso}}$ cannot become very large but for
some special constraints on the Bertrami parameters, for instance,
if \ $a_- \neq a_+$ \  and both \ $a_{\pm} \ll 1$ \ or the
condition (\ref{B19}) is satisfied.

\section{Conclusions and Summary}

In the present paper we derived {\it Quadruple} [{\it Triple}]
Beltrami relaxed states in e-p-i plasma with classical ions, and
degenerate electrons and positrons. Such a mix is often met in
both astrophysical and laboratory conditions.

The presence of the mobile ion component has a striking
qualitative effect; it converts, what would have been, a {\it
triple} Beltrami state to a new {\it quadruple} Beltrami state. In
the process, it adds  structures at a brand new macroscopic scale
\ $L_{\rm{macro}}$ \ (absent when ion concentration is zero) that
is much larger than the intrinsic skin depth ($\lambda =
c\sqrt{\frac{m^-}{8\pi n_0^- e^2}} $) of the lighter components.

Though primarily controlled by the mobile ion concentration, \
$L_{\rm{macro}}$ \ also takes cognizance of the electron and
positron inertias that could be considerably enhanced by
degeneracy. In fact even in the absence of ions
($L_{\rm{macro}}\rightarrow $ infinity), the Beltrami states could
be characterized by what could be called meso-scales -- the
temperature and degeneracy-boosted effective skin depths
$\lambda_{\rm{eff}}^{\pm}$ larger than $\lambda $ [according to
(6) \ ${\lambda_{\rm{eff}}}^{\pm}/\lambda = \sqrt{G_0^{\pm}} > 1$
and \ $1< \sqrt{G_0^{\pm}} < 5.6$ \ for densities \ $(10^{25} -
10^{32})\,cm^{-3}$] . At the same time it has to be emphasized
here that for larger scale to exist we do need an entirely
different mechanism -- a dynamic ion-species with a much lower
density and higher rest mass (justified by observations for many
astrophysical objects plasmas) -- this scale corresponds to the
ion skin depth enhanced, dramatically, by low density [$\lambda_i
= (\alpha \,m_-/m_i)^{-1/2}\ \lambda \gg \lambda $ ].

The creation of these new intermediate scales (between the system
size, and $\lambda$) adds immensely to the richness of the
structures that such an e-p-i plasma can sustain; many more
pathways become accessible for energy transformations. Such
pathways could help us better understand a host of quiescent as
well as explosive astrophysical phenomena - eruptions,
fast/transient outflow and jet formation, magnetic field
generation, structure formation, heating  etc. At the same time,
results found in present manuscript indicate that when the star
contracts, for example, its outer layers keep the multi-structure
character although density in the structures, as shown in
\citep{BSM_deg}, becomes defined by lighter components degeneracy
pressure. Future studies will include a detailed investigation of
present model to explore the evolution of multi-structure stellar
outer layers while contracting, cooling.

\section{Acknowledgements}

Authors acknowledge special debt to the Abdus Salam International
Centre for Theoretical Physics, Trieste, Italy. The work of S.M.M.
was supported by USDOE Contract No. DEFG 03-96ER-54366.


%



%

\appendix
\section{Appendix - Derivation of Quadruple Beltrami Equation}

\bigskip

The Ampere's law generally can be written in dimensionless
variables as:
\begin{equation}
\qquad \qquad  \nabla \times {\bf B} = \frac{1}{2}\left[ \alpha \
\frac{N_i}{N_{0i}}{\bf V}_i +(1-\alpha)\, \frac{N^+}{N_0^+}{\bf
V}_+ - \frac{N^-}{N_0^-}{\bf V}_-) \right] \label{B-Amp}
\end{equation}
and \\
\noindent (i) if \ $\alpha=1$ \ (e-i plasma,
quasineutralisty reads as $N_i=N^-=N$) \ we have:
\[
\qquad \qquad \qquad {\bf V}_-\equiv {\bf V}_e={\bf V}_i -
\frac{2}{N}\ \nabla \times {\bf B}
\]
leading to Double Beltrami (DB) states in e-i plasma with
degenerate electrons \citep{BSM_deg};

\noindent (ii) while when \ $\alpha=0$ \ (purely symmetric e-p
plasma, quasineutralisty reads as $N^+=N^-=N$) \ we have:
\[
\qquad \qquad \qquad {\bf V}_- - {\bf V}_+ = - \frac{2}{N}\ \nabla
\times {\bf B}
\]
that shall lead to higher Beltrami states when inertia effects in
electron and positron fluids are taken into account [similar
effect was discussed for relativistic non-degenerate plasmas in
\cite{iqbal-1,iqbal-3,dasgupta,Multi-B}].

Observations show that ion fluid fraction can be small ($\alpha
\ll 1$); also ion fluid velocity is much smaller than those for
lighter elements -- electron and positron fluids \ [${\bf V}_i \ll
{\bf V}_- , {\bf V}_+ $]  and, hence, one can imagine that the
mobility of ions can be ignored in most of the cases
\citep{relaxed} except when $\alpha = 1$ where, as it was shown in
\citep{mmns-1,mnsy,msms}, flow effects can be crucial in creating
the structural richness in astrophysical environments as well as
in the heating/cooling processes, Generalized Dynamo theory and
flow acceleration phenomena. The case of $\alpha =1$ - pure e-i
plasma with degenerate electrons was already studied in
\citep{BSM_deg} and it was shown that when ignoring inertia
effects in electron fluid the Double Beltrami states are
accessible in the system. Hence, it is expected, that when ion
fluid velocity is not negleted in e-p-i plasma with degenerate
electrons and positrons number of relaxed states can be either 2
(when ignoring degenerate {\it e(p)} fluids inertia effects
although $G^-\neq G^+$) or 4 (when degenerate fluids inertia
effects are taken into account); at the same time neglecting the
ion flow effects \ ($\alpha \to 0 , {\bf V}_i\to 0$) \ we shall
obtain the Single Beltrami state in former situation and the
Triple Beltrami States in latter case -- this problem is a scope
of our study below [see \cite{Multi-B} and its results].

\bigskip

Let us now show how Beltrami states may acquire new structures due
to degeneracy or/and the small fraction of mobile ions. We will
study an incompressible e-p-i plasma with the simplifying
assumption \ $\gamma_+ \sim\gamma_-\equiv 1$ \ that reduces the
Generalized Bernoulli Conditions to \ $G^+ + G^- = const$ \ . The
Ampere's law (\ref{B-Amp}), in dimensionless variables, is written
as
\begin{equation}
\qquad \qquad \nabla \times {\bf B} = \frac{1}{2}\,[\alpha \,{\bf
V}_i + (1-\alpha ){\bf V}_+ \ - \ {\bf V}_-] \ .
\label{B11}
\end{equation}
In terms of  the e-p plasma bulk flow average velocity
\begin{equation}
\qquad \qquad {\bf V} = \frac{1}{2}\,[(1-\alpha ){\bf V}_+ \ + \
{\bf V}_-] \label{B12'}
\end{equation}
\noindent one can express {\it the Generalized Momenta} for
positron and electron fluids as [$ {\bf
P}_{\pm}=G_0^{\pm}(n_0^{\pm})\,{\bf V}_{\pm} \ $] :
\begin{equation}
{\bf P}_{+} = \frac{G_0^+}{1-\alpha}\,({\bf V} \ + \ \nabla \times
{\bf B}-\frac{\alpha}{2}\,{\bf V}_i) \ ; \qquad \qquad \qquad {\bf
P}_{-} = G_0^-\,({\bf V} \ - \ \nabla \times {\bf
B}+\frac{\alpha}{2}\,{\bf V}_i) \ , \label{B13}
\end{equation}

\bigskip

Introducing \ $\beta \equiv \frac{G_0^-}{G_0^+} $ \ and using Eqs.
(\ref{B13}) in Eqs. (\ref{B9}), straightforward algebra leads to:
\[
\qquad \qquad {\bf V} = \eta \,\left( 2\beta \,G_0^+\,\nabla
\times \nabla \times {\bf B} - [a_+(1-\alpha)\beta - a_-]\,\nabla
\times {\bf B} \ + \ [1 + (1-\alpha)\,\beta ]\,{\bf B} \right) -
\]
\begin{equation}
\qquad \qquad \qquad \qquad \qquad \qquad - \ \alpha \,\beta \,
\nabla \times {\bf V}_i \ + \
\frac{\alpha}{2}\,[a_+(1-\alpha)\beta - a_-]\,{\bf V}_i
\label{B14}
\end{equation}
\[
{\rm with} \qquad \eta \,\equiv [a_+(1-\alpha )\beta + a_-]^{-1}
\quad .
\]

We have to add the Ion flow Beltrami condition ({\ref{IBC})
written for incompressible case as
\begin{equation}
\qquad \qquad \qquad {\bf B} + \zeta \,\nabla \times {\bf V}_i =
\alpha \,a_i \,{\bf V}_i \
\label{IBC-2}
\end{equation}
to close the system of equations for incompressible e-p-i
degenerate plasma.

\bigskip

Plugging the Eq. (\ref{B14}) into the Eq.-s (\ref{B13}) and then
using them in Eq.-s (\ref{B9}) we get
\begin{equation}
\qquad \qquad 2\beta \,(G_0^+)^2 \nabla \times \nabla \times
\nabla \times {\bf B} \ -\ 2G_0^+\ \alpha_1 \
\nabla\times\nabla\times {\bf B} \ + \ \alpha_2 \ \nabla \times
{\bf B} - \alpha_3 \ {\bf B} = \label{B-4}
\end{equation}
\[
\qquad \qquad \qquad \qquad \qquad = \alpha \,(G_0^+)^2\,\nabla
\times \nabla \times {\bf V}_i - \alpha\,(G_0^+) \alpha_1 \,
\nabla \times {\bf V}_i - \alpha\,(1-\alpha)\,a_+a_-\,{\bf V}_i \
,
\]
where
\begin{equation}
\qquad \qquad \alpha_1 = [a_+(1-\alpha)\beta - a_-] , \ \ \alpha_2
=[1 + (1-\alpha)\,\beta ]\ G_0^+ - 2\beta (1-\alpha)\ a_+\,a_- , \
\ \alpha_3=(1-\alpha)\ (a_+ + \ a_-) \ . \label{B-5}
\end{equation}
The equation (\ref{B-4}) for immobile ions (${\bf V}_i \equiv 0$)
will eventually give the so called ''Triple Beltrami'' equation
for the magnetic field \ ${\bf B}$ (i.e. l.h.s. of Eq.(\ref{B-4})
$ \equiv 0$).

\bigskip

Let us now simplify the equations when the ion density is just a
small fraction of the  density of the light species, i.e, \
$\alpha \ll 1 \Longrightarrow (1-\alpha) \to 1; \
[1+(1-\alpha)\beta \,] \to 2$ \ . After tedious but simple
algebra, one obtains, in this limit, the quadruple Beltrami
equation for ${\bf V}_i$:
\begin{equation}
\qquad (G_0^+)^2\,\nabla \times \nabla \times \nabla \times \nabla
\times {\bf V}_i - b_1\,(G_0^+)\,\nabla \times \nabla \times
\nabla \times {\bf V}_i + b_2\,(G_0^+)\,\nabla \times \nabla
\times {\bf V}_i - b_3\,\nabla \times {\bf V}_i + b_4\,{\bf V}_i =
0 \ , \label{QB-1}
\end{equation}
where
\begin{equation}
\qquad \qquad b_1 = (a_+ - a_-) + \frac{1}{2}\,a_i\,(G_0^+) ;
\qquad \qquad b_2 = [(G_0^+)-a_+a_-] +
\frac{\alpha}{\zeta}a_i\,(a_+ - a_-) +
\frac{\alpha}{\zeta}\,(G_0^+) ; \label{QB-2}
\end{equation}
\begin{equation}
\qquad \qquad b_3 = (a_+ - a_-) [1 -
\frac{\alpha}{\zeta}\,(G_0^+)] +
\frac{\alpha}{\zeta}\,a_i\,[(G_0^+)-a_+a_-] ; \qquad \qquad b_4 =
\frac{\alpha}{\zeta}\,[a_i\,(a_+ - a_-) -a_+a_-] . \label{QB-3}
\end{equation}
Notice, that in above relations the terms with coefficient \
$(\alpha/\zeta)$ \ will vanish for either \ $\alpha \to 0$ \ (when
there is no fraction of ions at all!) or \ $m_i\to \infty$ \
(which means that ions are immobile!). In such case we arrive to
Triple Beltrami Equations for \ ${\bf B}$ \ and \ ${\bf V}$ \ as
mentioned above. Also, it is interesting to note that due to
mobile ions (i.e. when \ $(\alpha/\zeta) \neq 0$) \ the large
scale is automatically there due to \ $b_4\neq 0$ \ in
Eq.(\ref{QB-1}). We will show this in detail in the Appendix B.

Solving the Eq.(\ref{QB-1}) for \ ${\bf V}_i$ \ and plugging it
into (\ref{IBC-2}) we will get the equation for \ ${\bf B}$ ; for
the pure incompressible e-p plasma it is better to use
Eq.(\ref{B14}) directly to find the magnetic field \ ${\bf B}$.

\section{Appendix - Scale separation in degenerate e-p-i Plasma
-- Quadruple Beltrami Structures}

Here we give some illustrative examples of Quadruple [Triple]
Beltrami states for degenerate e-p-i [e-p] plasmas interesting for
astrophysical context.

Introducing
\begin{equation}
\qquad \qquad \qquad b_1^{'} = \frac{b_1}{G_0^+} ; \qquad b_2^{'}
= \frac{b_2}{G_0^+} ; \qquad b_3^{'} = \frac{b_3}{(G_0^+)^2} ;
\qquad b_4^{'} = \frac{b_4}{(G_0^+)^2} \label{B-7}
\end{equation}
the Eq. (\ref{QB-1}) reads as:
\begin{equation}
\qquad \nabla \times \nabla \times \nabla \times \nabla \times
{\bf V}_i \ -\ b_1^{'} \,\nabla \times \nabla \times \nabla \times
{\bf V}_i \ + \ b_2^{'} \,\nabla \times \nabla \times {\bf V}_i \
 - \ b_3^{'} \,\nabla \times {\bf V}_i \ + \ b_4^{'}\,{\bf V}_i = \ 0 , \
\label{B-8}
\end{equation}

Equation (\ref{B-8}) can be written as
\begin{equation}
\qquad \qquad (curl - \mu_1)(curl - \mu_2)(curl - \mu_3)(curl -
\mu_4)\,{\bf V}_i = 0 \ ,
\label{B-9}
\end{equation}
where \ $\mu_k $-s \ define the coefficients in Eq. (\ref{B-8}) as
\[
\qquad \qquad b_1^{'}=\mu_1+\mu_2+\mu_3+\mu_4 \ , \qquad \qquad
b_2^{'}=\mu_1\,\mu_2 + \mu_1\,\mu_3 + \mu_1\,\mu_4 + \mu_2\,\mu_3
+ \mu_2\,\mu_4 + \mu_3\,\mu_4 \ ,
\]
\begin{equation}
\qquad \qquad b_3^{'}=\mu_1\,\mu_2\,\mu_3 + \mu_1\,\mu_2\,\mu_4 +
\mu_1\,\mu_3\,\mu_4 + \mu_2\,\mu_3\,\mu_4 \ , \qquad \qquad
b_4^{'} = \mu_1\,\mu_2\,\mu_3\,\mu_4. \label{B-10}
\end{equation}

The general solution of Eq.(\ref{B-9}) is a sum of four Beltrami
fields \ ${\bf F_k}$ \ (solutions of Beltrami Equations $\nabla
\times {\bf F}_k = \mu_k {\bf F}$) \ while eigenvalues ($\mu_k $)
of the {\it curl} operator are the solutions of the fourth order
equation
\begin{equation}
\qquad \qquad \mu^4-b_1^{'}\,\mu^3 + b_2^{'}\,\mu^2 - b_3^{'}\,\mu
+  b_4^{'}= 0 \ . \label{B-11}
\end{equation}

Since $b_4^{'} \to 0$ for our case of study -- e-p-i plasma with
small fraction of mobile ions: \ $\alpha \ll 1 ; \ m^- \ll m_i
\Longrightarrow $ \ that the Eq.(\ref{B-11}) can be reduced to
\begin{equation}
\qquad \qquad \mu\ (\mu - \mu_2)\ (\mu - \mu_3) \ (\mu - \mu_4)= 0
. \label{QB-ls}
\end{equation}
solving of which gives \ $\mu_1 \simeq 0$ \ -- hence, the large
scale structure existence is automatically guaranteed in such
plasma due to the presence of small fraction of mobile ions
(asymmetry due to small ion contamination); we stress here that
this scale is always there.

\bigskip

Now we will have to solve the remaining equation:
\begin{equation}
\qquad \qquad \mu^3-b_1^{'}\,\mu^2 + b_2^{'}\,\mu - b_3^{'}= 0 .
\label{TB-1}
\end{equation}

There are 3 possible simple, analytically tractable, scenarios:

\noindent (i) If
\begin{equation}
\qquad \qquad (a_+ - a_-) [1 - \frac{\alpha}{\zeta}\,(G_0^+)] = -
\frac{\alpha}{\zeta}\,a_i\,[(G_0^+)-a_+a_-]
\label{a}
\end{equation}
then \ $b_3^{'} \simeq 0$ , and eq. (\ref{TB-1}) reduces to:
\begin{equation}
\qquad \qquad \qquad \mu\ (\mu + a_1)\ (\mu - a_2) = 0  \ ,
\label{B-12}
\end{equation}
solving of which we find that \ $\mu_2 \simeq 0$ \ [another large
large scale now due to the asymmetry in degeneracy induced
inertias of electrons and positrons] is also guaranteed in such
plasma and two other short-scales are defined by Beltrami
Parameters \ $a_+$ \ and \ $a_-$ \ as: \ $\mu_3=-a_1 ; \
\mu_4=a_2$ \ , where
\begin{equation}
\qquad \qquad a_1 = \frac{1}{2}b_1^{'}\mp
\frac{1}{2}\sqrt{(b_1^{'})^2-4b_2^{'}} \qquad \qquad \rm{and}
\qquad \qquad a_2 = \frac{1}{2}b_1^{'} \pm
\frac{1}{2}\sqrt{(b_1^{'})^2-4b_2^{'}} \ .
\label{aa}
\end{equation}
In pure e-p plasma above conditions reduce to \ $a_+\sim a_-$ \
($a_1 = a_+$ \ while \ $a_2 = a_-$) and eventually there are only
3 scales in total (defining equation is {\it Triple Beltrami}).

\noindent (ii) If \ $\alpha $ \ and \ $\beta $ \ and other
parameters are such that both \ $b_1^{'} \simeq 0$ \ and \
$b_3^{'} \simeq 0$ , then \ $b_2^{'} = \mu_3 \mu_4 \neq 0$ . Here
again one of the roots is zero (let it be \ $\mu_2 = 0$ ), while
other two satisfy the relations:
\begin{equation}
\qquad \qquad \mu_3 +\mu_4 = 0  \ ; \qquad \mu_3\,\mu_4 = b_2^{'}
\ .
\label{B-13}
\end{equation}
Hence, also in this case length-scales are vastly separated (two
scales are of similar range (short scales) and one is
significantly large-scale): $|\mu_2|=|\mu_3|; \ \mu_2\simeq 0$ \ .
For such scenario Beltrami coefficients $a_{\pm}$ must be such
that \ $b_2^{'}<0$ \ is maintained; for pure e-p plasma this is
the case when \ $a_+ = a_- = a > (G/n)^{1/2}$ .

\noindent (iii) If \ $\alpha $ \ and \ $\beta $ \ and other
parameters are such that all \ $b_1^{'} \simeq 0$ , \ $b_2^{'}
\simeq 0$ \ and \ $b_3^{'} \simeq 0$ \ can be established in
addition to guaranteed \ $b_4^{'} \simeq 0$, then all the scale
parameters \ $\mu_1 , \mu_2 , \mu_3 , \mu_4$ \ become close to
zero -- no separation of length scales. For pure e-p plasma this
is the case when \ $a_+ = a_- = a = (G/n)^{1/2} $ .

Thus, we can conclude, that for a rather big range of parameters
there is a guaranteed scale separation in e-p-i plasma with
degenerate lighter components.

\vspace{0.8cm}

\section{Appendix - Scale separation in compressible degenerate e-p Plasma
-- Triple Beltrami Structures}

Note that with no fraction of ions \ [$\alpha =0 , \ \beta =1$ ] \
there is no charge separation and the scalar potential \ $\varphi
\equiv 0$ \ in Eq.-s (\ref{B6} - \ref{B7}); hence, $n^- = n^+ = n
$, \ $ {\bf B}=\nabla \times {\bf A} \ ; \ {\bf
E}=-\frac{1}{c}\,\frac{\partial}{\partial t}\,{\bf A} \ $ and the
assumption of \ $T_{\pm}\to 0$ \ leads to the initial temperature
symmetry between species; we have consequently \ $G^-=G^+=G(n)$ \
. The existence of soliton-like electromagnetic (EM) distributions
in such a fully degenerate electron-positron plasma was studied in
\cite{degenerate}.

Then, for pure compressible e-p plasma , if \ $\nabla
[G^{\pm}(n^{\pm})]$ \ is at a much slower rate than the spatial
derivatives of \ ${\bf B}$ \ and ${\bf V}_{\pm}$ , we can write
instead of (\ref{B14}) following relation [with corresponding \
$\eta = 1/(a_+ + a_- )$ ]:
\begin{equation}
\qquad \qquad {\bf V} = \eta \ \left( 2\,\frac{G}{n}\, \nabla
\times \frac{1}{n}\,\nabla \times {\bf B} - \frac{a_+ - a_-}{n}
\nabla \times {\bf B} \right) \ + \ \eta \frac{2{\bf B}}{n} \
\label{B15}
\end{equation}
and instead of (\ref{B-4}) we obtain:
\begin{equation}
\qquad \qquad
\nabla\times\left(\frac{G}{n}\right)\nabla\times\left(\frac{1}{n}\right)\nabla
\times {\bf B}
-\kappa_1\,\nabla\times\left(\frac{1}{n}\right)\nabla\times{\bf B}
+ \ \kappa_2 \,\nabla \times \left(\frac{G}{n}
-a_+\,a_-\right)\,{\bf B} \ - \ \kappa_3 \,{\bf B} \ = \ 0 , \
\label{B16}
\end{equation}
where \begin{equation} \qquad \qquad \qquad \kappa_1 = (a_+ -
a_-); \quad \kappa_2 = \frac{1}{G}; \quad \kappa_3 = \frac{a_+ -
a_-}{2G} \
\label{kappa}.
\end{equation}
After simple algebra we arrive to the defining equation for \
${\bf V} = \frac{1}{2}\,({\bf V}_+ \ + \ {\bf V}_-) $:
\begin{equation}
\qquad \qquad
\left(\frac{G}{n}\right)\nabla\times\left(\frac{1}{n}\right)\nabla\times\nabla\times{\bf
V} -\kappa_1
\,\left(\frac{1}{n}\right)\nabla\times\nabla\times{\bf V} \ +
\kappa_2 \,\nabla\times\left(\frac{G}{n}
-a_+\,a_-\right)\,\left(\frac{n}{G} \right)\,{\bf V} \ - \
\kappa_3 \,{\bf V} \ = \ 0 . \ \label{B17}
\end{equation}

Solution of Eq.-s (\ref{B16}) and (\ref{B17}) is possible
following the scenarios given in Appendix B, solutions will be
similar to those given after Eq.(\ref{TB-1}), just density
dependent. Estimation for the large scale \ $l_{\rm{meso}}$ \ in
case of pure degenerate e-p plasma, derived from the dominant
balance, gives:
\begin{equation}
\qquad \qquad \qquad l_{\rm{meso}}= \frac{|\kappa_2|}{|\kappa_3|}\
|(G/n)-a_+a_-| = 2\,\frac{|(G/n)-a_+a_-|}{|a_+ - a_-|} \gg 1
\label{B18}
\end{equation}
if
\begin{equation}
\qquad \qquad \qquad a_+ = a_- = a \neq
\left(\frac{G(n)}{n}\right)^{1/2} \ .
\label{B19}
\end{equation}
Hence, whenever the local density satisfies this condition there
is a guaranteed scale separation in the degenerate e-p plasma with
at least one large scale present.

\end{document}